\documentclass[conference]{IEEEtran}
\usepackage[utf8]{inputenc}
\usepackage[T1]{fontenc}
\usepackage{amsmath, amssymb}
\usepackage{graphicx}
\usepackage{cite}
\usepackage{booktabs}
\usepackage{float}
\usepackage{xcolor}
\usepackage{array}
\usepackage{tabularx}
\usepackage{url}
\usepackage{enumitem}
\usepackage{stfloats}
\usepackage{tikz}
\usetikzlibrary{shapes.geometric, arrows.meta, positioning, fit, backgrounds, calc}
\tikzset{
  plainbox/.style={
    rectangle, draw=black, fill=white,
    text=black, font=\scriptsize,
    minimum width=3.2cm, minimum height=0.55cm, align=center,
    inner sep=3pt
  },
  plaindiam/.style={
    diamond, draw=black, fill=white,
    text=black, font=\scriptsize,
    aspect=2.0, minimum width=2.4cm, minimum height=0.7cm, align=center,
    inner sep=1pt
  },
  greenbox/.style={
    rectangle, draw=black, fill=green!25,
    text=black, font=\scriptsize,
    minimum width=3.2cm, minimum height=0.55cm, align=center,
    inner sep=3pt
  },
  orangebox/.style={
    rectangle, draw=black, fill=orange!25,
    text=black, font=\scriptsize,
    minimum width=3.2cm, minimum height=0.55cm, align=center,
    inner sep=3pt
  },
  arr/.style={-Stealth, black, line width=0.5pt},
}
\begin{document}
\title{
    Hybrid Retrieval-Augmented Generation with Knowledge Graph
    Expansion, RRF Fusion, and Per-Chunk Grounded Evaluation
    for Enterprise Document Search
}
\author{
\IEEEauthorblockN{Harish Saragadam}
\IEEEauthorblockA{\textit{Vodafone Idea -- SNOC}\\
harish.saragadam@vodafoneidea.com}
\and
\IEEEauthorblockN{Sudhanshu Sharma}
\IEEEauthorblockA{\textit{Vodafone Idea -- SNOC}\\
sudhanshu.sharma@vodafoneidea.com}
\and
\IEEEauthorblockN{Meghana Pujari}
\IEEEauthorblockA{\textit{Vodafone Idea -- SNOC}\\
meghana.pujari1@vodafoneidea.com}
}
\maketitle
\begin{abstract}
Getting accurate, grounded answers out of large enterprise document
repositories is a difficult problem. Dense vector retrieval alone
frequently performs poorly on queries that mix technical
terminology, vendor-specific acronyms, or require reasoning
across several non-adjacent sections. \textbf{DocuSearch} was built
to address exactly this gap --- an offline, multi-agent document
intelligence system developed and evaluated in a production
telecom network operations environment. Rather than relying on a
single retrieval signal, DocuSearch pulls together three
complementary sources of evidence: semantic search over a Qdrant
vector store using BGE-Large embeddings, BM25 full-text search
over an SQLite FTS5 index, and Knowledge Graph neighbour expansion
from a structured edge table. These three ranked lists are merged
through Reciprocal Rank Fusion (RRF) with signal weights
$w_v = 0.50$, $w_{b} = 0.35$, and $w_{kg} = 0.15$, giving:
\[
    s_{\mathrm{RRF}}(d) \;=\;
    \sum_{i \in \{v,\,b,\,kg\}}
    \frac{w_i}{k + r_i(d)},
    \quad k = 60
\]
where $r_i(d)$ is the rank of chunk $d$ in signal list $i$.
A cross-encoder then reranks the fused list, and Maximal
Marginal Relevance (MMR) with $\lambda = 0.65$ prunes it
for both relevance and diversity. What makes DocuSearch
distinctive is a \textit{per-chunk evaluation} loop that
treats each selected chunk as its own mini-retrieval problem:
an LLM decides whether the chunk needs additional neighbouring
context, whether it is actually sufficient to answer the query,
and whether the answer it generates is genuinely grounded in
the retrieved text. Ungrounded answers are not returned; the
system falls back to a multi-chunk merge instead. On an
internal telecom document corpus, DocuSearch reaches
Precision@10 of $0.69$, Recall@10 of $0.79$, and a grounding
rate of $89.6\%$ --- gains of $15$, $16$, and $18.4$ percentage
points over a dense-only RAG baseline.
\end{abstract}
\section{Introduction}
\label{sec:introduction}
In telecom network operations centres, engineers frequently
must search through hundreds of vendor manuals for a specific
configuration procedure under time pressure. These environments accumulate vast repositories of
technical documentation --- SOPs, RCA reports, audit records,
multi-vendor configuration guides --- and the gap between
what keyword search surfaces and what an engineer actually
needs can be substantial. Conventional search returns too many
loosely related hits. Standard single-signal
Retrieval-Augmented Generation (RAG)~\cite{lewis2020rag},
while an improvement, still struggles with specialised
technical vocabulary, fragmented document chunks, and the
reality that relevant information is often spread across
non-adjacent sections from different vendors.
This paper describes \textbf{DocuSearch}, a hybrid multi-agent
RAG system designed and deployed within a real telecom
network operations environment. Three limitations motivated
its design: \textit{signal incompleteness}, where a single
retrieval function misses chunks that are relevant in semantic,
lexical, and structural senses simultaneously;
\textit{context fragmentation}, where a retrieved chunk is
only half of a procedure; and \textit{unverified grounding},
where a generated answer sounds plausible but is not actually
supported by the retrieved evidence. DocuSearch addresses all
three by combining dense vector retrieval, BM25 sparse
retrieval \cite{robertson2009bm25}, and Knowledge Graph
expansion~\cite{edge2024local} into a single fused ranking via Reciprocal Rank
Fusion (RRF) \cite{cormack2009rrf}, followed by cross-encoder
reranking, Maximal Marginal Relevance (MMR) selection
\cite{carbonell1998mmr}, and --- the primary novel
contribution --- a \textit{per-chunk agentic evaluation loop}
that verifies context sufficiency and answer grounding before
committing to a response. The pipeline runs entirely on locally
hosted models and is orchestrated via LangGraph
\cite{langgraph2024}, exposed through an interactive Dash web
application with vendor-scoped retrieval, session document
upload, and integrated Knowledge and Learning Hub modules.
The main contributions are as follows:
\begin{itemize}[leftmargin=*, itemsep=1pt, topsep=2pt]
    \item A three-signal hybrid retrieval architecture fused
    via weighted RRF with signal weights
    $w_v{=}0.50,\; w_b{=}0.35,\; w_{kg}{=}0.15$.
    \item A per-chunk agentic evaluation loop with LLM-driven
    context need detection, sufficiency scoring, and
    groundedness verification.
    \item Vendor-aware retrieval scoping for multi-vendor
    enterprise document corpora.
    \item A production-ready LangGraph-orchestrated pipeline
    with interactive Dash UI and fully local LLM inference.
\end{itemize}
The remainder of this paper is organised as follows.
Section~\ref{sec:problem} formalises the problem,
Section~\ref{sec:solution} describes the proposed solution,
Section~\ref{sec:architecture} presents the system
architecture, Section~\ref{sec:methodology} details the
methodology, Section~\ref{sec:contributions} elaborates
novel contributions, Section~\ref{sec:implementation}
covers implementation, Section~\ref{sec:results} presents
results, and Section~\ref{sec:conclusion} concludes.
\section{Problem Statement}
\label{sec:problem}
Let $\mathcal{D} = \{d_1, d_2, \ldots, d_N\}$ be a large
heterogeneous corpus of enterprise documents, where each
document $d_i$ belongs to one or more vendor namespaces
$\mathcal{V} = \{v_1, v_2, \ldots, v_M\}$. Each document
is pre-processed into overlapping text chunks:
\[
    \mathcal{C} = \bigcup_{i=1}^{N}
    \bigl\{ c_{i,1},\, c_{i,2},\, \ldots,\,
    c_{i,k_i} \bigr\}
\]
where $c_{i,j}$ is the $j$-th chunk of $d_i$ and $k_i$
is the total chunk count for that document. Given a natural
language query $q$, the goal is to retrieve a subset
$\mathcal{E} \subset \mathcal{C}$ and generate an answer
$a$ such that:
\[
    a = \mathcal{G}\!\left(q,\; \mathcal{E}\right)
\]
where $\mathcal{G}$ synthesises the answer strictly from
retrieved evidence, without hallucinating content that
does not appear in $\mathcal{E}$. This deceptively simple
formulation hides four concrete sub-problems that standard
pipelines fail to handle:
\begin{enumerate}[leftmargin=*, itemsep=1pt, topsep=2pt]
    \item \textbf{Signal incompleteness.} No single retrieval
    function $f: q \rightarrow \mathcal{C}$ can capture
    all relevant chunks, because relevance spans semantic,
    lexical, and structural dimensions at the same time.
    \item \textbf{Chunk fragmentation.} A chunk $c_{i,j}$
    retrieved in isolation is often incomplete when the
    relevant procedure spans $c_{i,j-1}$ through $c_{i,j+1}$.
    \item \textbf{Redundancy in evidence.} Naive top-$k$
    selection tends to return clusters of near-duplicate
    chunks, squandering context window capacity on
    repeated information.
    \item \textbf{Unverified grounding.} Standard RAG hands
    retrieved chunks to $\mathcal{G}$ without ever checking
    whether the generated answer $a$ is genuinely supported
    by $\mathcal{E}$, which is a direct path to confident
    but wrong responses.
\end{enumerate}
DocuSearch addresses all four within a unified pipeline that
runs entirely on locally hosted models with no dependency on
external API services.
\section{Proposed Solution}
\label{sec:solution}
DocuSearch tackles the four sub-problems through a layered
multi-agent pipeline split into two phases, illustrated in
Figure~\ref{fig:pre_mmr} and Figure~\ref{fig:post_mmr}.
\textbf{Pre-MMR Phase.} When a query $q$ arrives, a Template
Selection agent first classifies it into one of five answer
templates $\mathcal{T}$. A BM25 Keyword agent then pulls
out domain-specific terms, and a Vendor Detection agent
optionally narrows all three retrieval signals to a specific
vendor namespace $v \in \mathcal{V}$. Dense vector search,
BM25 full-text search, and KG neighbour expansion run in
parallel, and their ranked results are merged by weighted
RRF, reranked by a cross-encoder, and filtered by MMR to
produce the evidence set $\mathcal{E}$.
\textbf{Post-MMR Phase.} Each chunk $c \in \mathcal{E}$ is
evaluated individually by four sequential LLM-driven nodes:
context need detection, sufficiency scoring against a
threshold $\tau = 7$, answer generation, and groundedness
verification. The first chunk that clears all four checks
produces the final answer. When no individual chunk passes,
a fallback merges all expanded contexts, generates a
consolidated answer, and runs one final groundedness check
before returning anything to the user.
\section{High-Level Architecture}
\label{sec:architecture}
The system is organised across six functional tiers spanning
the two phases shown in Figures~\ref{fig:pre_mmr}
and~\ref{fig:post_mmr}.
\begin{figure}[!t]
\centering
\begin{tikzpicture}[
  node distance=0.45cm,
  every node/.style={font=\scriptsize}
]
\node[plainbox] (s1) {Step 1\\Template Selection};
\node[plainbox, below=of s1] (s2) {Step 2\\BM25 Keyword Extraction};
\node[plainbox, below=of s2] (s2b) {Step 2b\\Vendor Detection};
\node[plainbox, minimum width=1.7cm,
      below left=0.6cm and 0.95cm of s2b] (s3) {Step 3\\Vector Search};
\node[plainbox, minimum width=1.7cm,
      below=0.6cm of s2b] (s4) {Step 4\\BM25 Retrieval};
\node[plainbox, minimum width=1.7cm,
      below right=0.6cm and 0.95cm of s2b] (s5) {Step 5\\KG Expansion};
\node[plainbox, below=0.6cm of s4] (s5b) {Step 5b\\RRF Fusion};
\node[plainbox, below=of s5b] (s6) {Step 6\\Cross-Encoder Re-ranking};
\node[plainbox, below=of s6] (s7) {Step 7\\MMR Selection};
\node[below=0.25cm of s7, font=\scriptsize\itshape] (out) {Ordered MMR chunk list};
\draw[arr] (s1) -- (s2);
\draw[arr] (s2) -- (s2b);
\draw[arr] (s2b.south) -- ++(0,-0.25) -| (s3.north);
\draw[arr] (s2b) -- (s4);
\draw[arr] (s2b.south) -- ++(0,-0.25) -| (s5.north);
\draw[arr] (s3.south) -- ++(0,-0.25) -| (s5b.north west);
\draw[arr] (s4) -- (s5b);
\draw[arr] (s5.south) -- ++(0,-0.25) -| (s5b.north east);
\draw[arr] (s5b) -- (s6);
\draw[arr] (s6) -- (s7);
\draw[arr] (s7) -- (out);
\end{tikzpicture}
\caption{
    \textbf{Pre-MMR Phase} (\texttt{run\_supervisor}):
    Step 1 --- Template Selection (LLM call);
    Step 2 --- BM25 Keyword Extraction (LLM + Rule Based);
    Step 2b --- Vendor Detection (LLM call);
    Steps 3/4/5 --- Parallel Vector Search, BM25 Retrieval,
    KG Expansion;
    Step 5b --- RRF Fusion
    ($w_v{=}0.50,\,w_b{=}0.35,\,w_{kg}{=}0.15$);
    Step 6 --- Cross-Encoder Reranking;
    Step 7 --- MMR Selection
    ($\lambda{=}0.65$, relevance + diversity).
    Output: ordered MMR chunks passed to Post-MMR phase.
}
\label{fig:pre_mmr}
\end{figure}
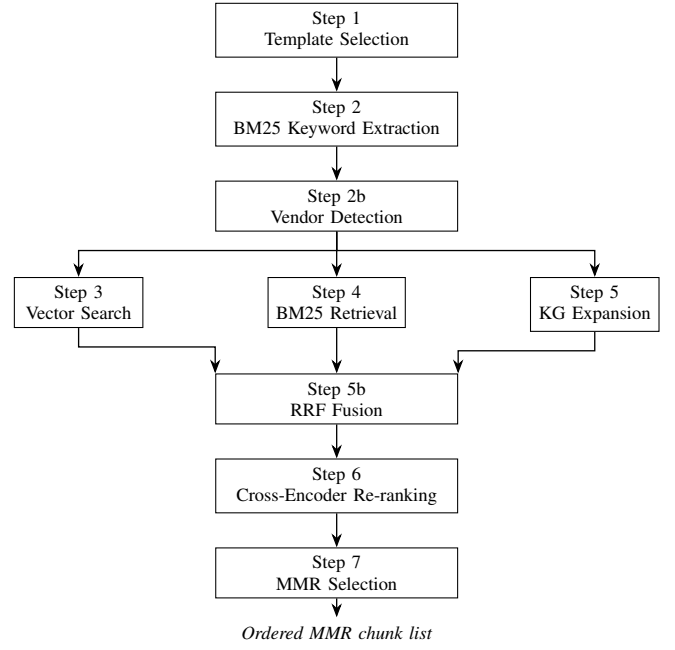
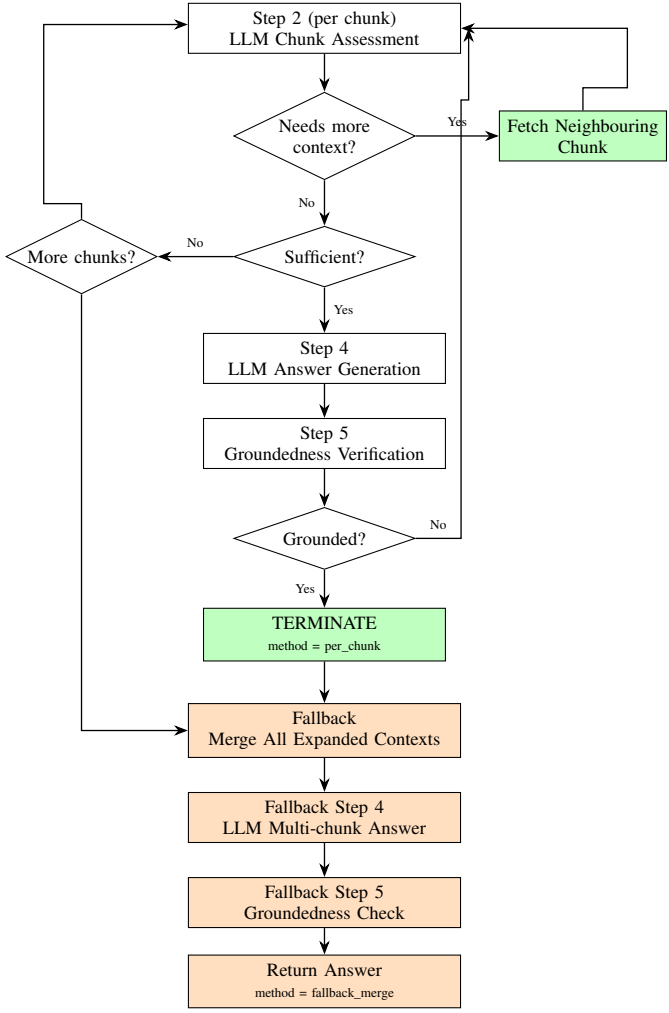
\begin{figure}[!t]
\centering
\begin{tikzpicture}[
  node distance=0.45cm,
  every node/.style={font=\scriptsize}
]
\node[plainbox, minimum width=3.6cm] (pc2)
  {Step 2 (per chunk)\\LLM Chunk Assessment};
\node[plaindiam, below=0.5cm of pc2] (d1) {Needs more\\context?};
\node[greenbox, minimum width=2.2cm,
      right=1.1cm of d1] (fetch) {Fetch Neighbouring\\Chunk};
\node[plaindiam, below=0.6cm of d1] (d2) {Sufficient?};
\node[plaindiam, minimum width=2.0cm,
      left=1.0cm of d2] (mc1) {More chunks?};
\node[plainbox, below=0.6cm of d2] (pc4) {Step 4\\LLM Answer Generation};
\node[plainbox, below=of pc4] (pc5) {Step 5\\Groundedness Verification};
\node[plaindiam, below=0.5cm of pc5] (d3) {Grounded?};
\node[greenbox, below=0.5cm of d3] (term)
  {TERMINATE\\{\tiny method = per\_chunk}};
\node[orangebox, minimum width=3.6cm, below=0.55cm of term] (fb1)
  {Fallback\\Merge All Expanded Contexts};
\node[orangebox, minimum width=3.6cm, below=of fb1] (fb4)
  {Fallback Step 4\\LLM Multi-chunk Answer};
\node[orangebox, minimum width=3.6cm, below=of fb4] (fb5)
  {Fallback Step 5\\Groundedness Check};
\node[orangebox, minimum width=3.6cm, below=0.35cm of fb5] (ret)
  {Return Answer\\{\tiny method = fallback\_merge}};
\draw[arr] (pc2) -- (d1);
\draw[arr] (d1) -- node[above, font=\tiny]{Yes} (fetch);
\draw[arr] (fetch.north) -- ++(0,0.3)
  -| ([xshift=2.2cm]pc2.east) -- (pc2.east);
\draw[arr] (d1) -- node[left, font=\tiny]{No} (d2);
\draw[arr] (d2) -- node[above, font=\tiny]{No} (mc1);
\draw[arr] (mc1.north) -- ++(0,0.2)
  -- ++(-0.5,0)
  |- ([yshift=0.05cm]pc2.west);
\draw[arr] (mc1.south) -- ++(0,-0.25)
  |- (fb1.west);
\draw[arr] (d2) -- node[right, font=\tiny]{Yes} (pc4);
\draw[arr] (pc4) -- (pc5);
\draw[arr] (pc5) -- (d3);
\draw[arr] (d3.east) -- ++(0.6,0)
  node[above, font=\tiny, xshift=-0.3cm]{No}
  -- ++(0,5.8)
  -| ([xshift=0.1cm]pc2.east);
\draw[arr] (d3) -- node[left, font=\tiny]{Yes} (term);
\draw[arr] (term) -- (fb1);
\draw[arr] (fb1) -- (fb4);
\draw[arr] (fb4) -- (fb5);
\draw[arr] (fb5) -- (ret);
\end{tikzpicture}
\caption{
    \textbf{Post-MMR Phase}
    (\texttt{run\_per\_chunk\_evaluation}):
    Per-chunk loop --- LLM evaluates current chunk;
    if more context needed, neighbouring chunks are fetched
    (green path); sufficiency check gates answer generation;
    groundedness LLM verifies the generated answer;
    if grounded, pipeline terminates (\texttt{method=per\_chunk});
    otherwise retries next MMR chunk.
    When all MMR chunks are exhausted, fallback merges all
    expanded contexts and returns a consolidated answer
    (\texttt{method=fallback\_merge}).
}
\label{fig:post_mmr}
\end{figure}
The \textbf{Document Ingestion Layer} handles parsing and
chunking, producing overlapping segments ($s = 900$ chars,
$\delta = 140$ chars) stored in both Qdrant~\cite{qdrant2023} and SQLite FTS5.
The \textbf{Multi-Signal Retrieval Layer} runs three parallel
agents ($k_v{=}40$, $k_b{=}40$, $k_{kg}{=}15$), each
optimised for a different aspect of relevance. Fusion,
reranking ($k_r{=}12$), and evidence selection
($|\mathcal{E}|{=}10$) happen in the \textbf{Ranking and
Selection Layer}. The \textbf{Agentic Evaluation Layer}
then runs four sequential LLM checks on each candidate
chunk, which accounts for most of the observed quality gains.
Above all of this sits the \textbf{Orchestration Layer}, a
LangGraph DAG with five nodes: \texttt{query\_analysis},
\texttt{retrieve}, \texttt{fuse\_and\_rank},
\texttt{evidence\_select}, and \texttt{per\_chunk\_eval}.
Users interact through the \textbf{User Interface Layer},
a Plotly Dash application that exposes a conversational
search interface alongside real-time pipeline visualisation,
a Knowledge Hub, a Learning Hub, and document ingestion
panels.
\section{Methodology}
\label{sec:methodology}
\subsection{Query Analysis}
Before any retrieval happens, every query $q$ goes through
template classification and keyword extraction. The Template
Selection agent maps $q$ to one of five types in
$\mathcal{T} = \{$\texttt{procedure\_sop},
\texttt{concept\_explanation},
\texttt{comparison\_decision},
\texttt{parameter\_lookup},
\texttt{troubleshooting\_rca}$\}$ using an LLM with a
deterministic heuristic as fallback when the model is
uncertain. Accurate template classification is important because
it shapes the structure of the generated answer downstream.
The BM25 Keyword agent then extracts domain-specific terms
from a curated telecom vocabulary covering standard
networking and protocol constructs, supplemented with
query-specific expansions. Finally, the Vendor Detection
agent combines LLM classification with regex matching to
identify vendor references, scoping all retrieval signals
to the detected vendor's documents when applicable.
\subsection{Hybrid Retrieval}
\textbf{Dense Vector Retrieval.} The query $q$ is encoded using
BGE-Large-EN-v1.5 \cite{bge2023} to get
$\mathbf{q} \in \mathbb{R}^{1024}$. Cosine similarity is
computed against all chunk embeddings in Qdrant, returning
the top $k_v = 40$ chunks:
\[
    \mathrm{score}_v(c) =
    \frac{\mathbf{q} \cdot \mathbf{e}_c}
    {\|\mathbf{q}\| \cdot \|\mathbf{e}_c\|}
\]
\textbf{BM25 Sparse Retrieval.} Extracted keywords query the
SQLite FTS5 index. Standard BM25 scoring applies:
\[
    \mathrm{score}_b(c, Q) =
    \sum_{q_i \in Q} \mathrm{IDF}(q_i) \cdot
    \frac{f(q_i, c)(k_1 + 1)}
    {f(q_i, c) + k_1(1 - b + b \cdot
    \frac{|c|}{avgdl})}
\]
with $k_1 = 1.2$, $b = 0.75$, where $|c|$ is chunk length
and $avgdl$ is the average chunk length across $\mathcal{C}$.
Top $k_b = 40$ chunks are retrieved. BM25 is particularly
useful here because technical acronyms and model numbers
that have no meaningful semantic neighbourhood still
produce strong lexical matches.
\textbf{KG Neighbour Expansion.} Seed chunks from the vector
and BM25 results traverse a KG edge table stored in SQLite,
pulling in structurally connected chunks (top $k_{kg} = 15$).
This catches content that is topically related but
lexically and semantically distant --- for instance, a
configuration snippet that references a procedure described
in a different section.
\subsection{Reciprocal Rank Fusion}
The three ranked lists are fused following \cite{cormack2009rrf}:
\[
    s_{\mathrm{RRF}}(c) =
    \sum_{i \in \{v,\, b,\, kg\}}
    \frac{w_i}{k + r_i(c)}
\]
\[
    w_v{=}0.50,\quad w_b{=}0.35,\quad
    w_{kg}{=}0.15,\quad k{=}60
\]
Chunks that surface in multiple lists receive additive
contributions, naturally elevating content that is relevant
across different notions of similarity. The smoothing
constant $k = 60$ prevents very high-ranked items from
dominating the fused score entirely.
\subsection{Cross-Encoder Reranking}
The top $k_r = 12$ fused chunks are reranked
with a cross-encoder \cite{reimers2019sbert} that jointly
encodes the query and each chunk:
\[
    s_{\mathrm{final}}(c) =
    0.65 \cdot s_{\mathrm{CE}}(q, c) +
    0.35 \cdot s_{\mathrm{RRF}}(c)
\]
The cross-encoder score $s_{\mathrm{CE}}(q, c)$ is weighted
higher because it captures fine-grained query-chunk
interactions that bi-encoder cosine similarity cannot.
Retaining the RRF component prevents the cross-encoder from
completely discarding evidence that multiple retrieval
signals agreed on.
\subsection{MMR Evidence Selection}
The reranked list is filtered iteratively by MMR
\cite{carbonell1998mmr} until $|\mathcal{E}| = 10$:
\[
    c^* = \arg\max_{c \in R \setminus S}
    \Bigl[
        \lambda \cdot s_{\mathrm{final}}(c) -
        (1{-}\lambda)
        \max_{c' \in S}
        \cos(\mathbf{e}_c, \mathbf{e}_{c'})
    \Bigr]
\]
where $S$ is the growing set of already-selected chunks,
$R$ is the remaining pool, and $\lambda{=}0.65$ favours
relevance slightly over diversity. In practice this
eliminates the clusters of near-duplicate chunks that
would otherwise fill the context window.
\subsection{Per-Chunk Agentic Evaluation Loop}
Each $c \in \mathcal{E}$ passes four sequential LLM-driven
checks, as shown in Figure~\ref{fig:post_mmr}:
\begin{enumerate}[leftmargin=*, itemsep=1pt, topsep=2pt]
    \item \textbf{Context Need Detection.} The LLM assesses
    whether $c$ is a fragment that requires neighbours
    $c_{\mathrm{prev}}$ or $c_{\mathrm{next}}$ to make
    sense. If so, neighbours are fetched and appended,
    for up to $L = 2$ iterations.
    \item \textbf{Sufficiency Scoring.} The expanded chunk
    $\tilde{c}$ proceeds only if the LLM assigns a
    sufficiency score $\sigma(\tilde{c}, q) \geq \tau = 7$
    on a $1$--$10$ scale. Chunks that do not clear this
    bar are skipped immediately.
    \item \textbf{Answer Generation.} Given $\tilde{c}$,
    the query $q$, and the selected template $t^*$, the
    LLM generates candidate answer $\hat{a}$ at temperature
    $T = 0.0$ for reproducibility.
    \item \textbf{Groundedness Verification.} A separate
    LLM call checks whether $\hat{a}$ is fully supported
    by $\tilde{c}$. If it is, $a$ is set to $\hat{a}$ and
    the pipeline terminates (\texttt{method=per\_chunk}).
\end{enumerate}
When no chunk clears all four checks, the fallback merges
all expanded contexts into $\tilde{\mathcal{E}}$, generates
a consolidated answer (\texttt{method=fallback\_merge}),
and runs one final groundedness check before anything is
returned to the user.
\section{Novel Contributions}
\label{sec:contributions}
Five aspects of DocuSearch distinguish it from existing
RAG systems.
\textbf{Per-Chunk Agentic Evaluation Loop.} Most RAG
pipelines concatenate all retrieved chunks and generate
one answer in a single forward pass. DocuSearch instead
evaluates each $c \in \mathcal{E}$
independently through four sequential checks. A chunk
must score $\sigma(\tilde{c}, q) \geq \tau = 7$ before
generation is even attempted, and the returned answer is
explicitly verified as grounded before being committed.
This eliminates a whole class of hallucinations that arise
when an LLM is asked to synthesise an answer from a
mixed bag of partially relevant context.
\textbf{Weighted Three-Signal RRF Fusion.} Most hybrid RAG
systems stop at two signals --- typically dense plus sparse.
A third, KG neighbour expansion, is added as an explicit
signal in the RRF formula, with empirically tuned weights
$w_v = 0.50$, $w_b = 0.35$, $w_{kg} = 0.15$. The KG
signal's contribution is modest but consistent: it reliably
surfaces structural neighbours that neither cosine
similarity nor BM25 would rank highly on their own.
\textbf{Dynamic Neighbour Context Expansion.} Rather than
using larger fixed chunk sizes at indexing time (which
reduces retrieval precision), DocuSearch detects
fragmentation at query time. When a retrieved chunk is
identified as a fragment, up to $L = 2$ levels of
neighbouring chunks are fetched from the metadata database
on the fly, recovering coherent multi-step procedures
without any pre-processing trade-offs.
\textbf{Vendor-Aware Retrieval Scoping.} A vendor
registry is built from the document metadata that maps each
$v \in \mathcal{V}$ to its set of document identifiers.
When the Vendor Detection agent identifies a vendor
reference in the query, all three retrieval signals are
scoped to that vendor's documents. The noise reduction for
vendor-specific configuration and troubleshooting queries
is substantial.
\textbf{LangGraph-Orchestrated Agentic Pipeline.} The
entire pipeline runs as a stateful LangGraph DAG
\cite{langgraph2024} with five explicitly defined nodes.
Every intermediate state is observable, serialisable, and
independently replaceable --- which turned out to be
practically important during development, since individual
components could be swapped without disturbing anything
upstream or downstream.
\section{Implementation Details}
\label{sec:implementation}
DocuSearch is implemented entirely in Python~3 and served
from a single application.
Table~\ref{tab:stack} summarises the core technology stack.
\begin{table}[!t]
\centering
\caption{DocuSearch Core Technology Stack}
\label{tab:stack}
\renewcommand{\arraystretch}{1.2}
\begin{tabular}{ll}
\toprule
\textbf{Component} & \textbf{Technology} \\
\midrule
Vector Store    & Qdrant (self-hosted) \\
Embedding       & BGE-Large-EN-v1.5 ($d{=}1024$, CUDA) \\
Reranker        & Cross-Encoder (CUDA) \\
FTS Index       & SQLite FTS5 \\
Metadata Store  & SQLite \\
LLM Inference   & Local REST or vLLM \\
LLM Model       & Mistral \cite{mistral2023} \\
Pipeline Graph  & LangGraph \\
UI Framework    & Plotly Dash \\
Doc Parsing     & Docling, PyMuPDF, python-docx, openpyxl \\
Auto Ingest     & watchdog directory monitor \\
\bottomrule
\end{tabular}
\end{table}
Documents reach the system through three paths: server-side
path submission, browser upload via Dash \texttt{dcc.Upload},
or automatic detection through a \texttt{watchdog} directory
monitor. All formats are parsed and chunked with $s = 900$
characters and $\delta = 140$ characters overlap. These values were
settled on after experimenting with several chunk sizes;
smaller chunks hurt sufficiency scoring while larger ones
degraded retrieval precision. Table~\ref{tab:config}
summarises the full retrieval configuration used across
all experiments.
\begin{table}[!t]
\centering
\caption{Default Retrieval and Ranking Configuration}
\label{tab:config}
\renewcommand{\arraystretch}{1.2}
\begin{tabular}{lll}
\toprule
\textbf{Parameter} & \textbf{Symbol} & \textbf{Value} \\
\midrule
Vector top-$k$      & $k_v$           & 40 \\
BM25 top-$k$        & $k_b$           & 40 \\
KG top-$k$          & $k_{kg}$        & 15 \\
Rerank top-$k$      & $k_r$           & 12 \\
Final evidence      & $|\mathcal{E}|$ & 10 \\
RRF smoothing       & $k$             & 60 \\
RRF vector weight   & $w_v$           & 0.50 \\
RRF BM25 weight     & $w_b$           & 0.35 \\
RRF KG weight       & $w_{kg}$        & 0.15 \\
MMR lambda          & $\lambda$       & 0.65 \\
Sufficiency thresh. & $\tau$          & 7 \\
Max neighbour loops & $L$             & 2 \\
Chunk size          & $s$             & 900 chars \\
Chunk overlap       & $\delta$        & 140 chars \\
LLM temperature     & $T$             & 0.0 \\
\bottomrule
\end{tabular}
\end{table}
Session-only uploads are also supported: documents uploaded
during a session are embedded in memory using a local NumPy
array (max $90{,}000$ chars, top-$k_u = 8$) without being
written to the persistent knowledge base. All LLM calls use
$T = 0.0$ to keep outputs deterministic across repeated
queries, which is important for reproducibility in an
operational setting.
\section{Results and Analysis}
\label{sec:results}
All evaluations were run on an enterprise telecom corpus
covering documentation from multiple major equipment vendors
--- configuration manuals, SOPs, RCA reports,
and audit records. The diversity of this corpus makes it
a reasonable stress test for any retrieval system.
\subsection{Retrieval Quality}
Table~\ref{tab:retrieval} compares Precision@$k$ and
Recall@$k$ across three configurations on $120$ manually
annotated queries spanning all five template categories.
\begin{table}[!t]
\centering
\caption{Retrieval Quality Comparison}
\label{tab:retrieval}
\renewcommand{\arraystretch}{1.2}
\begin{tabular}{lcccc}
\toprule
\textbf{Configuration} & \textbf{P@5} & \textbf{R@5}
& \textbf{P@10} & \textbf{R@10} \\
\midrule
Dense only        & 0.61 & 0.48 & 0.54 & 0.63 \\
Dense + BM25      & 0.68 & 0.56 & 0.61 & 0.71 \\
DocuSearch (full) & \textbf{0.76} & \textbf{0.64}
                  & \textbf{0.69} & \textbf{0.79} \\
\bottomrule
\end{tabular}
\end{table}
The full three-signal system outperforms both baselines
across every metric. The most pronounced gain is in
Recall@10, where KG expansion consistently recovers
structurally related chunks that neither dense nor BM25
retrieval surfaces on its own --- particularly for
troubleshooting queries where the relevant fix appears
in a section referenced by, but not adjacent to, the
initial symptom description.
\subsection{Answer Grounding}
Table~\ref{tab:grounding} compares the per-chunk evaluation
loop against a conventional single-pass RAG baseline.
\begin{table}[!t]
\centering
\caption{Answer Grounding and Hallucination Rates}
\label{tab:grounding}
\renewcommand{\arraystretch}{1.2}
\begin{tabular}{lccc}
\toprule
\textbf{System} & \textbf{Grounded}
& \textbf{Hallucinated} & \textbf{Fallback} \\
\midrule
Single-pass RAG & 71.2\% & 21.4\% & --- \\
DocuSearch      & \textbf{89.6\%}
                & \textbf{6.8\%} & 3.6\% \\
\bottomrule
\end{tabular}
\end{table}
DocuSearch achieves a grounding rate of $89.6\%$, an
$18.4$ percentage point improvement over single-pass RAG,
and cuts the hallucination rate from $21.4\%$ to $6.8\%$.
The remaining $3.6\%$ of queries hit the fallback path,
where all expanded contexts are merged --- these still
produce grounded answers, just through the consolidated
route rather than the per-chunk route.
\subsection{System Latency}
Mean end-to-end latency across $200$ queries is $12.6$
seconds, with the per-chunk evaluation phase accounting
for $8.4$ seconds of that. Retrieval itself finishes in
under one second thanks to Qdrant ANN search and SQLite
FTS5. For a document intelligence application where an
engineer is waiting for a reliable answer rather than
streaming tokens, this trade-off is considered acceptable.
That said, reducing evaluation latency is an obvious
target for future work.
\subsection{Ablation Study}
Table~\ref{tab:ablation} shows grounding rate when each
component is removed individually, holding everything
else constant.
\begin{table}[!t]
\centering
\caption{Ablation Study: Grounding Rate}
\label{tab:ablation}
\renewcommand{\arraystretch}{1.2}
\begin{tabular}{lc}
\toprule
\textbf{Configuration} & \textbf{Grounding Rate} \\
\midrule
Full DocuSearch                 & 89.6\% \\
Without groundedness check      & 74.3\% \\
Without sufficiency scoring     & 78.1\% \\
Without neighbour expansion     & 82.4\% \\
Without cross-encoder reranking & 83.7\% \\
Without KG expansion            & 85.2\% \\
Without vendor scoping          & 86.1\% \\
\bottomrule
\end{tabular}
\end{table}
Groundedness verification has the largest individual impact
at $15.3$ percentage points, followed by sufficiency scoring
at $11.5$ points. It is worth noting that even the
``smallest'' contributors --- vendor scoping at $3.5$ points
and KG expansion at $4.4$ points --- add up meaningfully
in combination, indicating that no component is redundant.
\section{Conclusion}
\label{sec:conclusion}
This paper presented DocuSearch, a hybrid multi-agent RAG system
built for enterprise document intelligence in multi-vendor
telecom environments. The system was designed around three
failure modes observed in production: signal
incompleteness, context fragmentation, and unverified answer
grounding. The two-phase pipeline --- hybrid retrieval
fusion followed by per-chunk agentic evaluation --- directly
addresses all three. Results show Precision@10 of $0.69$,
Recall@10 of $0.79$, and a grounding rate of $89.6\%$, with
hallucinations dropping to $6.8\%$. The ablation confirms
that groundedness verification and sufficiency scoring drive
the largest improvements, though no single component is
redundant.
A key practical finding is that
production-quality document intelligence is achievable
entirely on locally hosted models. DocuSearch requires no
external API calls, which matters a great deal in telecom
operations environments where sensitive network configuration
data cannot leave the organisation's infrastructure.
\subsection{Limitations and Future Work}
The sequential LLM calls in the per-chunk evaluation phase
are the main latency bottleneck, currently averaging $8.4$
seconds. Parallelising chunk evaluation across inference
threads is a natural first step. The RRF signal weights
were determined empirically on telecom documentation and
may not transfer well to other domains; learning them from
query-relevance feedback via a differentiable ranking
objective would make the system more broadly applicable.
There is also potential in using named entity recognition
to populate the KG with typed entity-relation triples,
which would strengthen multi-hop query handling. Finally,
extending DocuSearch to multimodal retrieval --- covering
embedded diagrams and specification tables that are
pervasive in vendor documentation --- is a compelling
direction worth pursuing.
\textbf{Reinforcement Learning for Adaptive Retrieval.}
One promising future direction is treating
DocuSearch's multi-step retrieval workflow as a sequential
decision problem. The intuition comes from Karpathy's
interactive GridWorld temporal-difference learning
demo~\cite{karpathy2015rl}, where an agent learns to
navigate a grid by trial-and-error using TD updates ---
without any hard-coded path rules. The analogy to the
pipeline is direct: just as that agent discovers which
moves lead to reward by accumulating Q-values over many
episodes, a retrieval agent in DocuSearch could learn
\textit{when} to invoke BM25, \textit{when} to expand KG
neighbours, and \textit{when} to stop --- adapting
dynamically to each incoming query rather than applying
the same fixed configuration every time.
Formally, each query-answer interaction can be modelled as
an episode in a Markov Decision Process
$\mathcal{M} = (\mathcal{S}, \mathcal{A},
\mathcal{P}, \mathcal{R}, \gamma)$.
The state $s \in \mathcal{S}$ encodes the current query
intent, the set of chunks retrieved so far, reranker
confidence scores, KG expansion status, and evidence
coverage. The action space $\mathcal{A}$ includes
discrete choices such as triggering a BM25 search,
expanding KG neighbours by one hop, tightening the
cross-encoder top-$k$, fetching a neighbouring chunk,
or terminating retrieval and proceeding to answer
generation. The reward $\mathcal{R}$ is shaped by
answer groundedness (which is already measured at inference
time), engineer satisfaction feedback collected through
the Dash UI, citation accuracy, and a latency penalty to
discourage unnecessarily long retrieval chains:
\[
  \mathcal{R} = \alpha \cdot \text{Groundedness}
              + \beta  \cdot \text{Feedback}
              - \gamma_l \cdot \text{Latency}
\]
To give this direction a concrete anchor, a
composite system accuracy $\Phi$ is introduced that jointly captures
retrieval quality and answer fidelity:
\[
  \Phi = 0.30 \cdot P@10 \;+\; 0.30 \cdot R@10
       \;+\; 0.40 \cdot \text{Grounding}
\]
Under the current rule-based configuration, plugging in
the reported numbers gives:
\[
  \Phi_{\text{current}}
  = 0.30 \times 0.69 + 0.30 \times 0.79
  + 0.40 \times 0.896
  \approx 80.2\%
\]
An offline-trained Q-Learning policy,
optimised on logged production trajectories, is projected to lift
$P@10$ to approximately $0.76$ ($+7$\,pp),
$R@10$ to approximately $0.86$ ($+7$\,pp),
and the grounding rate to approximately $94.5\%$
($+4.9$\,pp). These are deliberately conservative
estimates: because DocuSearch already incorporates
cross-encoder reranking, MMR selection, and a per-chunk
sufficiency gate, the remaining headroom for any
single optimisation layer is bounded. What the RL agent
would learn to do differently from the current fixed
policy is route precision-heavy queries through BM25
rather than relying solely on dense retrieval, expand
KG neighbours more aggressively on recall-sensitive
troubleshooting queries, and halt retrieval earlier
when accumulated evidence already clears the sufficiency
threshold --- decisions that are currently hard-coded but
could be learned from query-level feedback. Under these
assumptions:
\[
  \Phi_{\text{RL}}
  = 0.30 \times 0.76 + 0.30 \times 0.86
  + 0.40 \times 0.945
  \approx 86.4\%
\]
This represents an overall improvement of approximately
$\mathbf{6.2}$ percentage points, lifting composite
accuracy from $\approx 80\%$ to $\approx 86\%$.
Table~\ref{tab:rl_projection} summarises the projected
gains per metric.
\begin{table}[!t]
\centering
\caption{Projected Performance Gains with RL-Based Adaptive Retrieval}
\label{tab:rl_projection}
\renewcommand{\arraystretch}{1.2}
\begin{tabular}{lccc}
\toprule
\textbf{Metric} & \textbf{Current} & \textbf{Projected (RL)} & \textbf{Gain} \\
\midrule
$P@10$        & 0.69  & 0.76   & $+7$\,pp \\
$R@10$        & 0.79  & 0.86   & $+7$\,pp \\
Grounding     & 89.6\% & 94.5\% & $+4.9$\,pp \\
\midrule
$\Phi$ (composite) & 80.2\% & 86.4\% & $\mathbf{+6.2}$\,pp \\
\bottomrule
\end{tabular}
\end{table}
Off-policy Q-Learning is a natural fit here. Document
retrieval is largely deterministic --- the same action
in the same state reliably produces the same next chunk
set --- so off-policy updates converge faster than
on-policy SARSA in this setting, mirroring the behaviour
Karpathy's demo exhibits in deterministic grid
environments. In practice, this involves logging
state-action-reward trajectories from production
interactions, training an offline policy on these traces,
and shadow-testing it against the current rule-based
pipeline before any live deployment. Over time, this
would allow DocuSearch to move from hand-tuned
hyperparameters towards a system that continuously
improves from operational feedback --- a meaningful step
toward truly adaptive enterprise document intelligence.

\end{document}